# Hierarchical Scaling in Systems of Natural Cities

Yanguang Chen[1], Bin Jiang[2]

(1. Department of Geography, College of Urban and Environmental Sciences, Peking University, Beijing 100871, P.R. China. E-mail: chenyg@pku.edu.cn; 2. Faculty of Engineering and Sustainable Development, Division of GIScience, University of Gävle, SE-801 76 Gävle, Sweden. Email: bin.jiang@hig.se )

**Abstract**: Hierarchies can be modeled by a set of exponential functions, from which we can derive a set of power laws indicative of scaling. The solution to a scaling relation equation is always a power law. The scaling laws are followed by many natural and social phenomena such as cities, earthquakes, and rivers. This paper is devoted to revealing the power law behaviors in systems of natural cities by reconstructing the hierarchy with cascade structure. The cities of America, Britain, France, and Germany are taken as examples to make empirical analyses. The hierarchical scaling relations can be well fitted to the data points within the scaling ranges of the size and area of the natural cities. The size-number and area-number scaling exponents are close to 1, and the allometric scaling exponent is slightly less than 1. The results show that natural cities follow hierarchical scaling laws and hierarchical conservation law very well. The hierarchical scaling law proved to be derived from entropy maximization principle, and this suggests that the evolution of natural cities is dominated by entropy maximization laws. This study is helpful for scientists to understand the power law behavior in the development of cities and systems of cities.

## 1. Introduction

Hierarchy is one of the basic characters of complex systems such as cities and networks of cities. A hierarchy can be mathematically described with a power law or a pair of exponential laws. Recently years many scientists are interested in hierarchical structure of natural and social systems (Pumain, 2006). A fractal object is a self-similar hierarchy (Frankhauser, 1998; Mandelbrot, 1982).

According to the ideas from fractal cities, a city or a system of cities can be treated as a hierarchy with cascade structure (Batty and Longley, 1994; Frankhauser, 1994). A finding is that a self-similar hierarchy can be described with two or three exponential functions, from which it follows a set of power functions indicative of scaling (Chen, 2008; Chen, 2012). These exponential and power functions are not only the portrayals of the city. They were found years ago and can be employed to describe a number of physical and social systems, e.g., rivers (Horton, 1945; Schumm, 1956; Strahler, 1952), seismicity (Gutenberg and Richter, 1954), coronary arteries (Chen, 2015a; Jiang and He, 1990; Jiang and He, 1990), systems of cities (Chen, 2008; Chen, 2012). What is more, the two sets of functions can be used to characterize fractal structure and the bifurcation process indicating the route from periodic oscillation to chaos (Chen, 2008). This suggests that the exponential and power equations represent universal laws which reflect a kind of ubiquitous order. The power functions can be regarded as a group of hierarchical scaling laws followed by both social systems such as cities and natural systems such as rivers.

Scaling suggests some type of scale invariance in a process of continuous transformation. In other words, to scale (contract or dilate) something by a constant factor does not change the spatial or temporal property of thing (Mandelbrot, 1982; Williams, 1997). The idea from scaling is very important to model scale-free phenomena. More and more scientists become aware of the importance of scaling analysis in urban studies (e.g. Batty, 2008; Batty and Longley, 1994; Bettencourt, 2013; Bettencourt *et al*, 2007; Chen, 2008; Frankhauser, 1998; Jiang, 2013; Lobo *et al*, 2013; Rybski *et al*, 2009). Meanwhile, many puzzling issues arise from the research on scaling of cities (Arcaute *et al*, 2015; Louf and Barthelemy, 2014a). Many questions are still pending and require much more studies before finding satisfying answers to them. Anyway, scaling laws often reveal the general principles underlying the structure of a physical problem (West *et al*, 2002). Scaling analysis is an effective approach to urban spatio-temporal and hierarchical modeling. Scaling relations take on power laws, and a power law can be decomposed into two exponential laws based on hierarchical structure. Exponential laws can be derived from the principle of entropy maximization, and this indicates that power laws and thus scaling are involved with entropy maximization processes (Chen, 2012).

The hierarchical scaling laws are associated with many mathematical laws of cities. The models are found and reconstructed by Chen (2008), who once explored the relationships between Zipf's law (Zipf, 1949), Christaller's central place hierarchy (Christaller, 1933/66), Beckmann's city hierarchy model (Beckmann, 1958), Davis' $2^n$ rule (Davis, 1978), and Berry-Woldenberg's analogy between rivers and central places (Woldenberg and Berry, 1967). Hierarchical scaling analysis can be employed to de-noise city rank-size distributions and reveal the regularity of urban evolution. This paper is devoted to revealing and describing the deep structure of systems of natural cities using the hierarchical scaling relations. We agree with Pumain (2006) who once argued that the analysis of the hierarchical organization of complex systems such as cities can provide new insight for understanding systems' evolution and emergence of order. The rest parts of this paper are organized as follows. In Section 2, the mathematical expressions of exponential laws and power laws for hierarchical structure are illuminated. In Section 3, three hierarchical scaling laws are applied to the natural cities of America, Britain, France, and Germany, and the results are illustrated. In Section 4, the main points are summarized, and the principle of entropy maximization is employ to explain the power law behaviors of natural cities. Finally, we reach the chief conclusions of this study.

## 2. Mathematical models

### 2.1 Hierarchical exponential laws

The urban hierarchy with cascade structure can be described from two complementary angles of view. The longitudinal distributions can be described with exponential functions, and the latitudinal relationships can be described with power functions. Considering a geographical region with $n$ cities, we can organize the cities into a hierarchy comprising $M$ classes according to the generalized $2^n$ rule (Chen, 2012; Jiang and Yao, 2010). Based on the top-down order, the cascade structure of the urban hierarchy can be modeled by a set of exponential functions such as

$$N_m = N_1 r_n^{m-1}, \quad (1)$$

$$S_m = S_1 r_s^{1-m}, \quad (2)$$

$$A_m = A_1 r_a^{1-m}, \quad (3)$$

where $m$ refers to the order number of city class ($m$=1, 2, …, $M$), $N_m$ denotes the number of cities of order $m$, correspondingly, $S_m$ and $A_m$ represent the mean population size and urban area at the $m$th class. The parameters are as below: $N_1$ denotes the number of the top-class cities, $S_1$ and $A_1$ are the mean size and area of the first-class cities, $r_n=N_{m+1}/N_m$ refers to the interclass **number ratio** of cities, $r_s=S_m/S_{m+1}$ to the city **size ratio**, and $r_a=A_m/A_{m+1}$ to the urban **area ratio**. Generally speaking, $N_1=1$, but for the three-parameter Zipf's distribution, $N_1>1$ (Chen, 2016). Equations (1), (2) and (3) compose the mathematical expressions of the generalized $2^n$ principle (Chen, 2012), which is based on Beckmann-Davis models (Beckmann, 1958; Davis, 1978). The first equation represents the *number law*, the second equation represents the *size law*, and the third equation represents the *area law* of urban hierarchies (Chen, 2008). The three exponential laws can be derived by using the method of entropy maximizing (Chen, 2012). For a self-similar hierarchy, if $r_n=2$ as given, then it will follow that $r_s \rightarrow 2$, and if $r_s=2$ as given, then it will follow that $r_n \rightarrow 2$, where the arrow denotes "approach". If $r_n=r_s=2$, the generalized $2^n$ principle will return to the standard $2^n$ principle. In this instance, we will have $T_m=N_m S_m=N_1 S_1=S_1$, where denotes the total population at the $m$th level. This implies a property of hierarchical conservation of size distributions.

A set of Zipf's models is hidden behind the three exponential laws. From equations (1), (2), (3) we can derive Zipf's laws of population size distribution and area size distribution. Where population size distribution is concerned, three types of Zipf's models can be derived. If $r_n=r_s$, we can derive an one-parameter Zipf's model, $S(k)=S_1/k$, where $k$ refers to rank, and $S_1$ is a parameter indicative of the largest city (Gabaix, 1999a; Gabaix, 1999b). The one-parameter Zipf's model is termed pure Zipf's law in literature (Batty, 2006). If $r_n \neq r_s$, we can derive a two-parameter Zipf's model, $S(k)=S_1/k^q$, where $q$ is the second parameter indicating scaling exponent. If $r_n \neq r_s$ and the largest city cannot influence the whole geographical region, we can derive a three-parameter Zipf's model, $S(k)=C/(k+h)^q$, where $h$ is the third parameter indicating translational factor, and $C$ denotes proportionality coefficient (Chen, 2016). This suggests that Zipf's distribution can act as an indication of self-similar hierarchy.

## 2.2 Hierarchical power laws

The relationships between exponential laws and power laws suggest the relationships between simplicity and complexity. Exponential laws indicate growth, distribution, and process with characteristic scales, while power laws indicate allometry, fractal, and pattern without characteristic scales. The former suggests simplicity, and the latter suggests complexity. The exponential laws and power laws can be integrated into the same framework with hierarchical scaling concept. This suggests a hidden link between simplicity and complexity. From the above exponential laws, it follows a set of power-law models as follows

$$N_m = \mu S_m^{-D}, \quad (4)$$

$$N_m = \eta A_m^{-d}, \quad (5)$$

$$A_m = a S_m^b, \quad (6)$$

where the parameters can be expressed as $\mu = N_1 P_1^D$, $D = \ln r_n / \ln r_s$, $\eta = N_1 A_1^d$, $d = \ln r_n / \ln r_a$, $a = A_1 P_1^{-b}$, and $b = \ln r_a / \ln r_s$. Among these parameters, $D$ and $d$ denote the fractal dimension of city size distributions, and is the allometric scaling exponent of urban hierarchy. In fact, $b$ the is the ratio of the fractal dimension $D$ to the dimension $d$, that is, $b = D/d = (\ln r_n / \ln r_s)/(\ln r_n / \ln r_a) = \ln r_a / \ln r_s$. Apparently, from equations (1) and (2), we can derive equation (4); from equations (1) and (3), we can derive equation (5); from equations (2) and (3), or from equations (4) and (5), we can derive equation (6). This implies that, for the cascade structure of a hierarchy of cities, exponential laws and power laws represent two different sides of the same coin. The exponential laws can be directly derived from the principle of entropy maximization, and thus entropy maximization can be employed to indirectly explain the power laws of cities.

The hierarchical scaling of cities performs power law behaviors and can be expressed with the three power functions. Equation (4) suggests the *size-number scaling* in a hierarchy of cities. It is equivalent to the Pareto law of population-size distribution, and $D$ is the fractal dimension of urban hierarchies measured with city size such as population. Equation (5) suggests the *area-number scaling* of cities. It is equivalent to the Pareto law of city-area distribution, and $d$ can be regarded as the fractal dimension of urban hierarchies measured with urbanized area. Equation (6) suggests the hierarchical allometric scaling relation between urban area and size, and $b$ is the allometric scaling exponent of urban hierarchy. The inverse functions of equations (4) and (5) are equivalent to the Zipf's laws of population size distribution and area size distribution. This implies that Zipf's distribution is just a signature of hierarchical scaling. In scientific research, one of difficult problems of mathematical modeling rests with spatial dimension (Waldrop, 1992). Hierarchy and network represent two different sides of the same coin (Batty and Longley, 1994). Network structure is associated with spatial recursive subdivision (Goodchild and Mark, 1988).

Using hierarchical scaling, we are able to find new way of modeling spatial distribution and network organization (Figure 1).

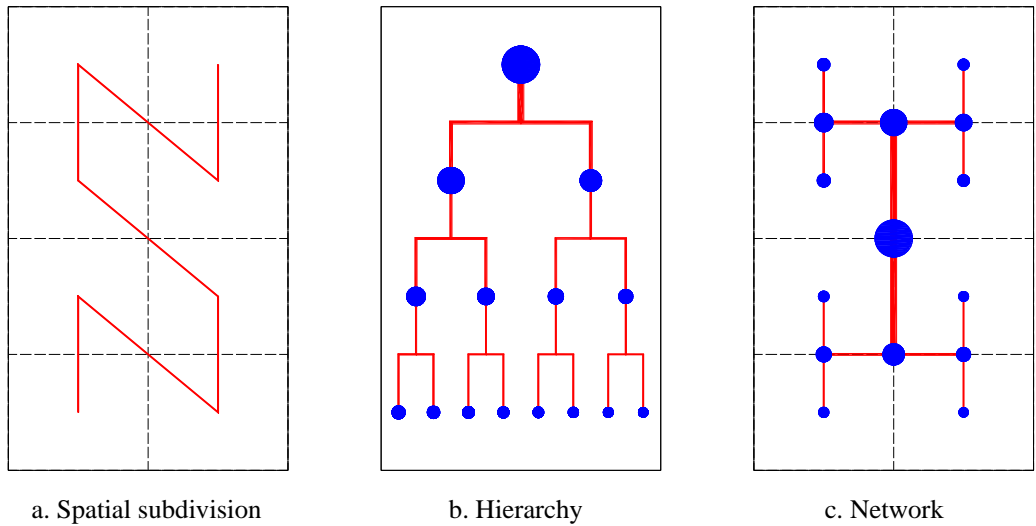

a. Spatial subdivision  b. Hierarchy  c. Network

**Figure 1 Spatial recursive subdivision, hierarchy, and network structure of cities**

[**Note**: The rank-size distribution of cities can be organized into a self-similar hierarchy, which corresponds to a cascade network. The network structure is based on strict recursive subdivision of geographical space. See Batty and Longley (1994) and Goodchild and Mark (1987)]

## 3. Empirical analysis

### 3.1 Study area and methods

The validity and rationality of the mathematical models can be verified and evaluated through empirical observation data. In fact, the success of natural sciences just rests heavily with their great emphasis on the role of interplay between quantifiable data and models (Louf and Barthelemy, 2014a). Four systems of cities in Europe and America can be employed to testify the hierarchical scaling laws and the related models about cities. Jiang and his coworkers (Jiang and Jia, 2011; Jiang and Liu, 2012) proposed a concept of *natural city* and developed a new approach to measuring objective city sizes using street nodes or blocks. Urban block is an ordinary concept, and the street nodes are defined as street intersections and ends. Using an identification algorithm of urban boundary, Jiang's research group was able to delineate boundaries of natural cities and yield city areal extents. To a degree, this method bears an analogy with the city clustering algorithm (CCA) proposed by Rozenfeld *et al* (2008, 2011). In theoretical geography, an urban boundary is termed *urban envelope*, and the city areal extent is called *urban area* (Batty and Longley, 1994; Longley *et al*, 1991). Thus urban area can be determined by a city's areal extent containing a large number of street blocks or nodes. The number of street nodes is significantly correlated with the population size of cities. The city data are extracted from massive volunteered geographic information OpenStreetMap databases through some data-intensive computing processes, and four datasets on the cities of **America (USA)**, **Britain** (UK), **France**, and **Germany** are available.

The empirical analysis can start from investigating Zipf's distribution, which can be regarded as a signature of the hierarchies with cascade structure. If cities in a region follow Zipf's law, they can be organized into a self-similar hierarchy (Chen, 2012). On the other hand, if a system of cities possesses cascade structure, the cities in the system will follow Zipf's law. It has been shown that the cities in the four countries follow Zipf's law (Jiang and Jia, 2011; Jiang and Liu, 2012). Applying the generalized $2^n$ rule to the above-mentioned datasets, we can create four self-similar hierarchies of European and American cities. Suppose that these systems of cities follow the abovementioned pure Zipf's law. Then the cities in each country can be reorganized into a hierarchy with cascade structure. Table 1 is presented for understanding the operational process of hierarchical reconstruction.

**Table 1 A standard hierarchy with cascade structure based on the pure rank-size distribution of cities (the first four classes)**

| Level $m$ | Hierarchical reconstruction of the rank-size distribution ($S_m=\ln2/2^{m-1}$) | | | | | | | |
|---|---|---|---|---|---|---|---|---|
| 1 | $S_1=1$ | | | | | | | |
| 2 | $S_2=1/2$ | | | | $S_3=1/3$ | | | |
| 3 | $S_4=1/4$ | | $S_5=1/5$ | | $S_6=1/6$ | | $S_7=1/7$ | |
| 4 | $S_8=1/8$ | $S_9=1/9$ | $S_{10}=1/10$ | $S_{11}=1/11$ | $S_{12}=1/12$ | $S_{13}=1/13$ | $S_{14}=1/14$ | $S_{15}=1/15$ |
| … | … | … | … | … | … | … | … | … |

**Note**: The theoretical foundation was given by Chen (2012).

Several algorithms can be adopted to evaluate the scaling exponents. The most common ones include the least squares method (LSM) (Chen, 2015b), maximum likelihood method (MLM) (Clauset *et al*, 2009; Newman, 2005), and major axis method (MAM) (Chen, 2016; Zhang and Yu, 2010). Recent years, the MLM is often used to identify power-law distributions, and it is treated as the most available approach to estimating power exponents. In fact, the power-law relations of this work are based on exponential functions, and are converted into logarithmic linear models. It was demonstrated that if the observations come from an exponential family and mild conditions are satisfied, the least-squares estimates are identical to the maximum-likelihood estimates (Charnes *et al*, 1976). What is more, if the errors of a linear model belong to the normal distribution, the least squares estimators are also identical to the maximum likelihood estimators. All in all, the function of an algorithm is to estimate the parameter values of a mathematical model rather than judge the form of a model's expression. Any algorithm has its advantages and disadvantages, sphere of application, and applicative conditions. The precondition of applying the MLM to observational data is that the variables satisfy the joint normal distribution. Unfortunately, for human systems such as cities, the observational data do not always meet the joint normal distribution. In this case, the LSM is an advisable approach to estimating power exponent values (Chen, 2015b). The models' parameters are evaluated by using the least squares calculations.

**3.2 Results and findings**

The systems of cities in USA, UK, France, and Germany can be well described with hierarchical scaling formulae. In light of the generalized $2^n$ principle expressed by equations (1) and (2), we can organize the cities in each country into a hierarchy with cascade structure. The city number in the $m$th level is $N_m$=1, 2, 4, …, $2^{m-1}$,…. The level numbers of urban hierarchies in the four countries are 15, 11, 11, and 13, respectively. The last levels are lame-duck classes because that city numbers are not big enough. Based on the hierarchical structure, we can calculate the average city size $P_m$ and the corresponding average urban area $A_m$ at each level (Table 2). The city numbers in different classes are designed according to the $2^n$ rule and satisfy equation (1). It is easy to testify that city size $P_m$ and urban area $A_m$ follow exponential distribution and meet equations (2) and (3), respectively, but the lame-duck classes are two outliers due to lack of adequate cities. Strictly speaking, the first class is usually an outlier because the largest city is often an exception (Chen, 2012). In fact, a mathematical law always becomes ineffective when the scale of measurement is too large or too small.

**Table 2 The reconstructed hierarchical systems of natural cities with cascade structure for America, Britain, France, and Germany (2010)**

| Class | America | | | Britain | | | France | | | Germany | | |
|---|---|---|---|---|---|---|---|---|---|---|---|---|
| $m$ | $N_m$ | $S_m$ | $A_m$ | $N_m$ | $S_m$ | $A_m$ | $N_m$ | $S_m$ | $A_m$ | $N_m$ | $S_m$ | $A_m$ |
| 1 | 1 | 290503.000 | 1194500.000 | 1 | 46299.000 | 91938.879 | 1 | 62242.000 | 133817.492 | 1 | 28866.000 | 40265.780 |
| 2 | 2 | 213517.000 | 783975.000 | 2 | 10993.500 | 20368.164 | 2 | 10877.000 | 21812.770 | 2 | 25354.500 | 36563.584 |
| 3 | 4 | 176132.500 | 746975.000 | 4 | 5230.250 | 8434.331 | 4 | 6972.250 | 17203.731 | 4 | 19394.000 | 25766.545 |
| 4 | 8 | 115663.500 | 501678.125 | 8 | 3946.375 | 6340.649 | 8 | 3541.875 | 9044.170 | 8 | 10758.875 | 12169.475 |
| 5 | 16 | 60697.125 | 236468.750 | 16 | 2034.188 | 2925.685 | 16 | 2097.688 | 5529.493 | 16 | 5168.750 | 6245.420 |
| 6 | 32 | 31127.938 | 134110.156 | 32 | 1059.219 | 1802.168 | 32 | 1179.563 | 3175.526 | 32 | 2528.500 | 2940.365 |
| 7 | 64 | 15077.375 | 71724.609 | 64 | 530.453 | 1000.051 | 64 | 483.063 | 1622.074 | 64 | 1131.203 | 1541.896 |
| 8 | 128 | 7804.250 | 36437.695 | 128 | 246.094 | 457.709 | 128 | 220.945 | 728.457 | 128 | 588.867 | 836.570 |
| 9 | 256 | 3992.852 | 19124.316 | 256 | 96.258 | 204.449 | 256 | 105.547 | 319.176 | 256 | 309.762 | 455.393 |
| 10 | 512 | 2068.379 | 10039.502 | 512 | 38.986 | 77.708 | 512 | 44.010 | 102.194 | 512 | 164.701 | 248.410 |
| 11 | 1024 | 1072.855 | 5235.742 | *228* | 21.311 | 19.792 | *217* | 24.249 | 22.879 | 1024 | 82.616 | 128.013 |
| 12 | 2048 | 560.370 | 2922.583 | | | | | | | 2048 | 36.726 | 55.571 |
| 13 | 4096 | 288.579 | 1593.188 | | | | | | | *1065* | 20.488 | 18.542 |
| 14 | 8192 | 145.798 | 903.534 | | | | | | | | | |
| 15 | *14922* | 75.202 | 530.333 | | | | | | | | | |

**Note**: The original city datasets of America (USA), Britain (UK), France, and Germany is available from: http://fromto.hig.se/~bjg/scalingdata/ .

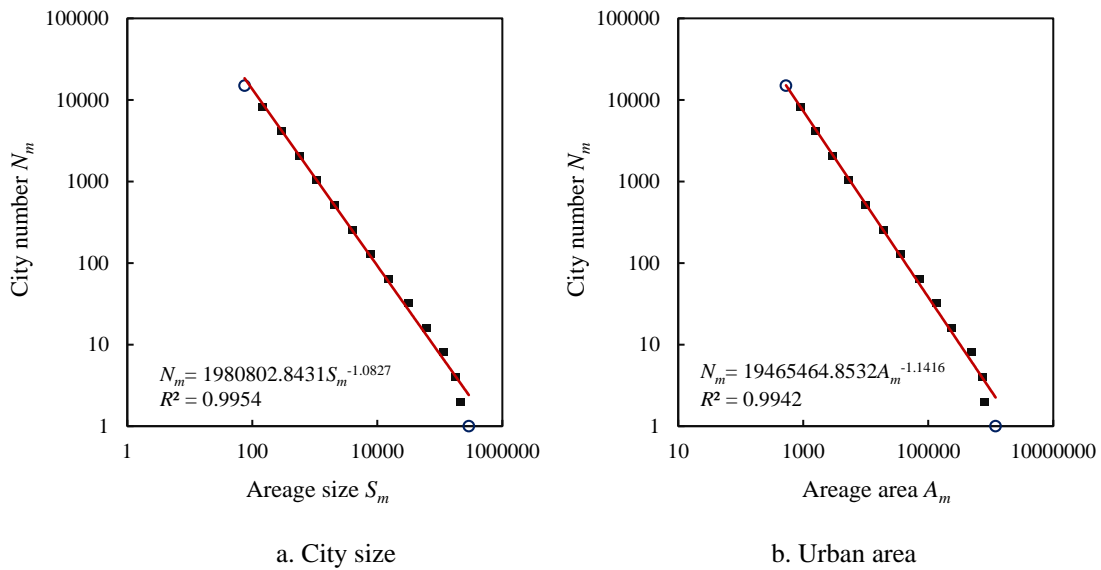

a. City size  b. Urban area

**Figure 2 The hierarchical scaling relationships between size (block/street node quantity) and area (physical extent) of American cities**

[**Note:** The small circles represent top classes and the lame-duck classes, respectively. Removing the first and last classes yields a scaling range. The slopes based on the scaling ranges indicate the fractal parameters of city size and area distributions. The ratio of the size dimension $D$ to the area dimension $d$ is close to the allometric scaling exponent $b$, i.e., $b \approx D/d$. Similarly hereinafter.]

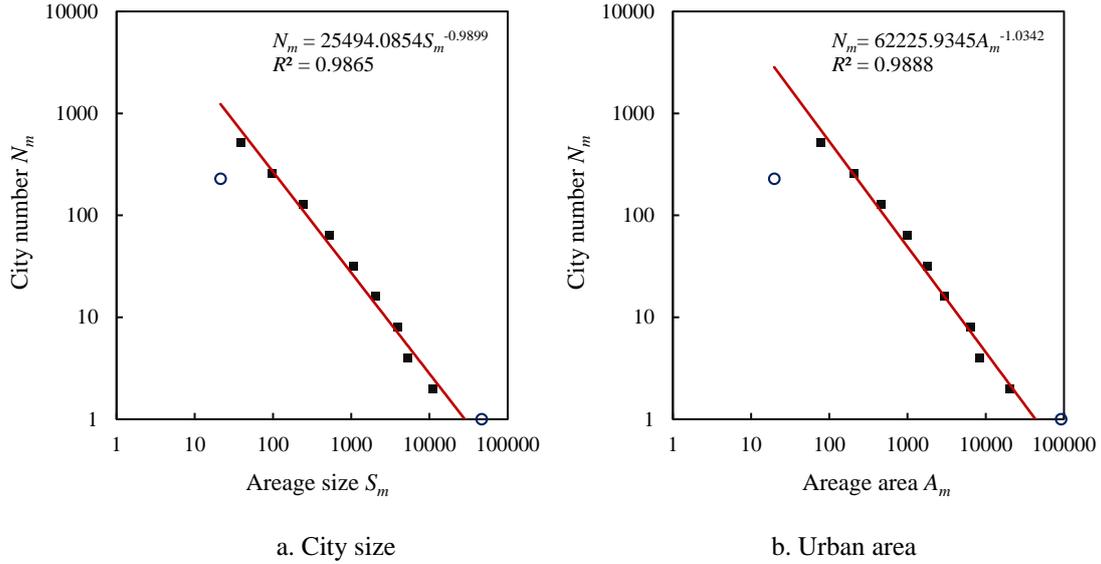

**Figure 3 The hierarchical scaling relationships between size (block/street node quantity) and area (physical extent) of British cities**

The exponential distributions of city size and urban area result in the power-law relations between city number, size, and area. The exceptional values in the exponential laws often manifest themselves on the log-log plots for power laws. In fact, if the scale is too large or too small, a power-law relation always breaks down (Bak, 1996; Chen, 2008). Thus the extreme classes always form exceptional points, and there exists a scaling range between the two extremes. For American cities, the last class of cities is out of trend lines and forms outliers, but the first class of cities is normal (Figure 2). For the Britain, French, and German cities, both the first and last classes are exceptional values (Figure 3, Figure 4, and Figure 5). For comparability, the first class of American hierarchy of cities is treated as an outlier, which does not influence the results and conclusions significantly. Removing the first and last data points as outliers yields scaling ranges for the relations between city number and city size or urban area. All the data points within the scaling range follow power law and take on double logarithmic linear relationships. In short, without considering the first and last classes, the relation between city size and number can be described with equation (4), and the relation between urban area and number can be described with equation (5). Fitting equations (4) and (5) to the datasets in Table 2, we can evaluate the parameters by the least squares calculation. The scaling exponent values are close to 1, and the $d$ value (area exponent) is slightly greater than the $D$ values (size exponent). The ratio of $D$ to $d$ can be termed fractal dimension quotient of urban hierarchies. As indicated above, if $D$ approaches 1, the total "population" of the $m$th level approaches a constant $S_1$. Despite the fact that there are always many smaller cities than larger ones (Batty, 2006; Jiang and Jia, 2012; Jiang and Yin, 2014), the product of average size and city number at each class seems to be invariable.

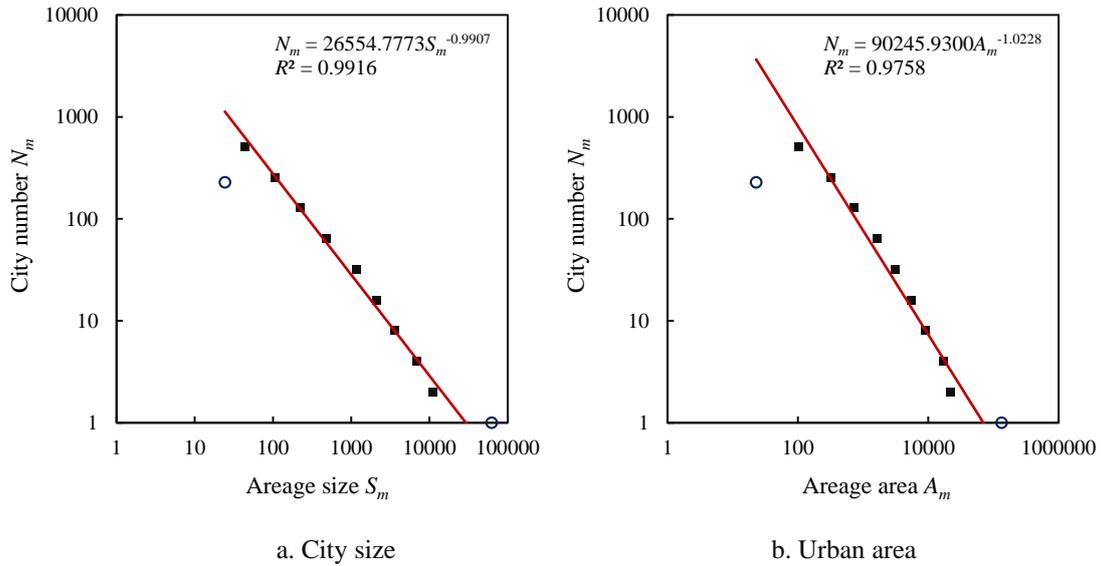

a. City size

b. Urban area

**Figure 4 The hierarchical scaling relationships between size (block/street node quantity) and area (physical extent) of French cities**

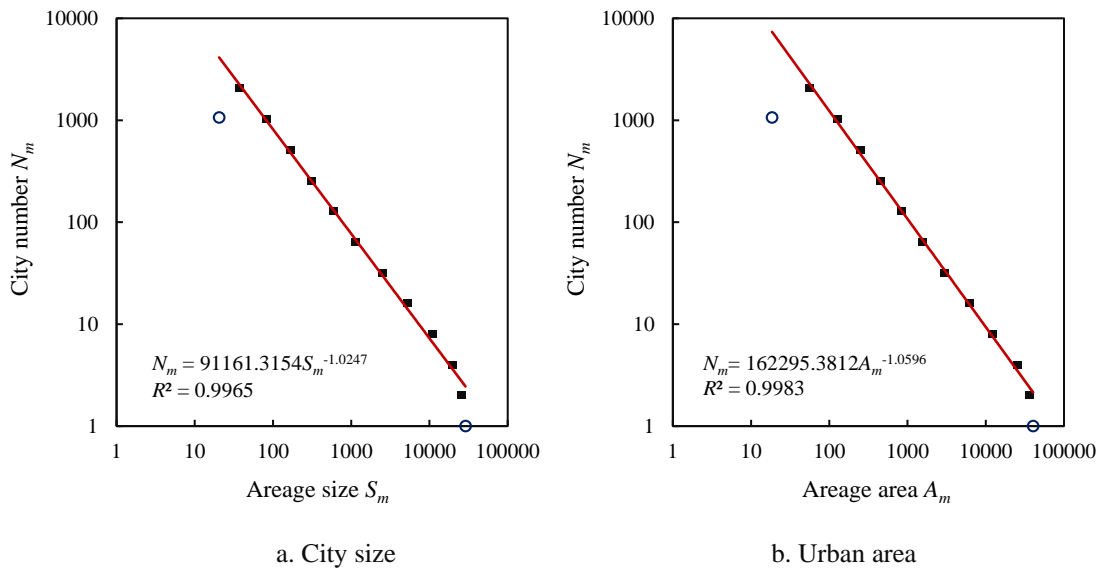

a. City size

b. Urban area

**Figure 5 The hierarchical scaling relationships between size (block/street node quantity) and area (physical extent) of German cities**

The relationships between city number and city size or urban area are a pair of fractal dimension relations, from which it follows an allometric scaling relation between city size and urban area. Using the data displayed in Table 1, we can estimate the allometric scaling exponent values. Corresponding to the exponential models and fractal models above mentioned, the first and last classes are treated as outliers so that the allometric parameters and fractal parameters are more comparable with one another. The allometric scaling of the hierarchies of cities in the four European and American countries is clear and convincing. For American cities, all the data points follow the allometric scaling law; for the cities of UK, French, and Germany, the last levels, i.e., the lame-duck classes, are exceptional points (Figure 6). The main results are shown in Table 3, in which we can see the way and effect of data processing.

The four study areas, USA, UK, French, and Germany, are all developed countries, and the levels of urbanization are near the capacity values. The allometric scaling properties of these urban hierarchies are as below: First, the allometric scaling exponent is close to but less than 1. This suggests that the relative growth rate of city size is slightly less than that of urban area. Second, the allometric scaling exponent is equivalent to the fractal dimension quotient. In theory, the allometric exponent is the ratio of the fractal dimension of urban population size distribution to that of urban area-size distribution. Where empirical analysis is concerned, the allometric exponent is close to the fractal dimension ratio. Generally speaking, for the developing systems of cities, the fractal dimension of population-size distribution is significantly less than that of area-size distribution. The allometric scaling exponent values come between 2/3 and 1, and always approach to 0.85 (Chen, 2008; Chen, 2010; Louf and Barthelemy, 2014b). However, for the developed urban systems, the difference between the two types of size distribution dimension is not significant. Therefore, the allometric scaling exponent is close to 1. Otherwise, a system will lose its balance (Bertalanffy, 1968).

**Table 3 The allometric scaling exponents and related parameters and statistics of four self-similar hierarchies of Euramerican natural cities (2010)**

| Type | Parameter and statistic | America | Britain | France | Germany |
|---|---|---|---|---|---|
| Size distribution | Fractal dimension ($D$) | 1.0827 | 0.9899 | 0.9907 | 1.0247 |
| | Standard error ($\sigma$) | 0.0222 | 0.0438 | 0.0344 | 0.0203 |
| | Goodness of fit ($R^2$) | 0.9954 | 0.9865 | 0.9916 | 0.9965 |
| Area distribution | Fractal dimension ($d$) | 1.1416 | 1.0342 | 1.0228 | 1.0596 |
| | Standard error ($\sigma$) | 0.0263 | 0.0415 | 0.0609 | 0.0147 |
| | Goodness of fit ($R^2$) | 0.9942 | 0.9888 | 0.9758 | 0.9983 |
| Size-area allometry | Allometric exponent ($b$) | 0.9476 | 0.9571 | 0.9578 | 0.9672 |
| | Standard error ($\sigma$) | 0.0063 | 0.0179 | 0.0289 | 0.0124 |
| | Goodness of fit ($R^2$) | 0.9995 | 0.9976 | 0.9937 | 0.9985 |
| Fractal dimension quotient | $D/d$ | 0.9484 | 0.9571 | 0.9686 | 0.9671 |
| Related quantity | City number ($n$) | 31305 | 1251 | 1240 | 5160 |
| | Level number ($M$) | 15 | 11 | 11 | 13 |
| | Scaling range | 2~14 | 2~10 | 2~10 | 2~12 |
| | Degree of freedom | 11 | 7 | 7 | 9 |

**Note:** For significance level $\alpha=0.01$ and degree of freedom $df=7$, the threshold value of Pearson correlation coefficient is $R_{0.01, 7}=0.7977$. The minimum correlation coefficient values of the four cases is $R=0.9968$.

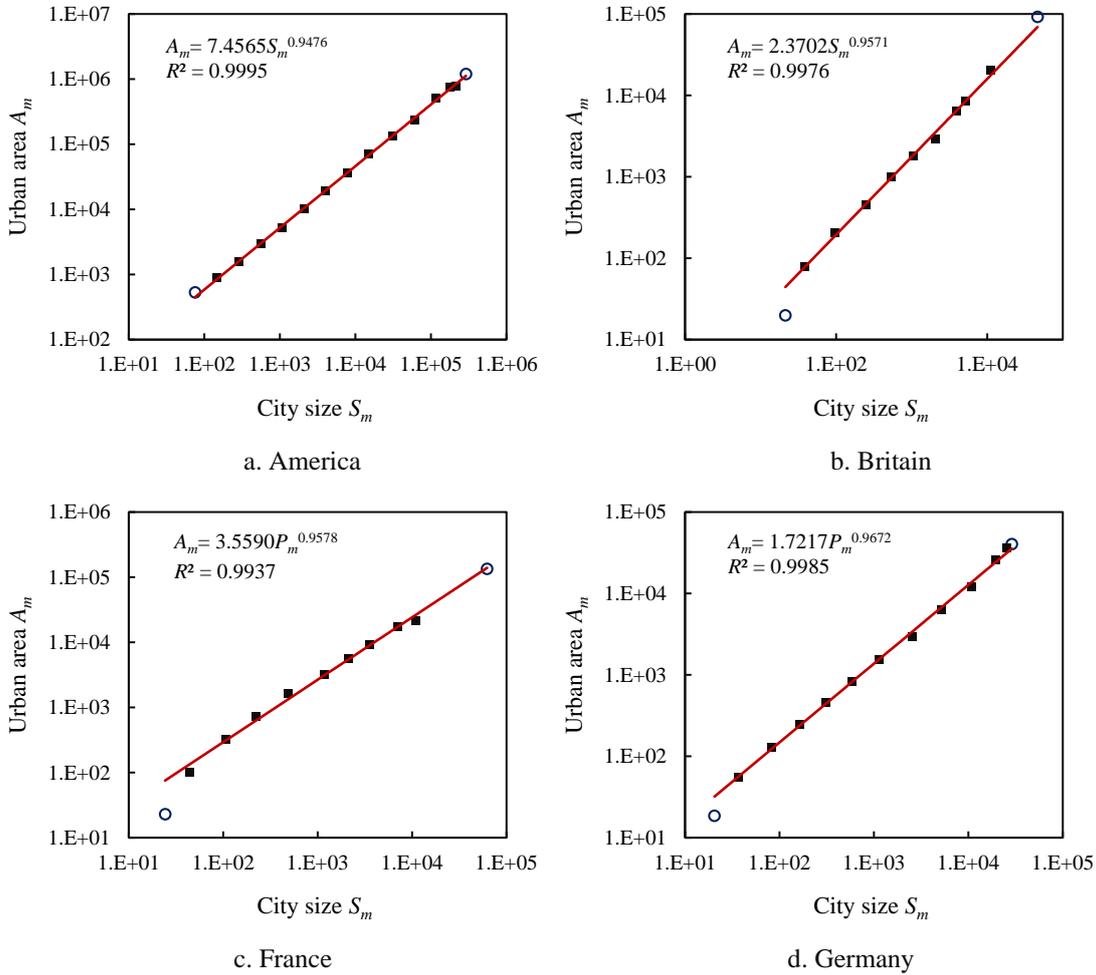

a. America     b. Britain

c. France     d. Germany

**Figure 6 The hierarchical allometric scaling patterns of four systems of natural cities (America, Britain, France, and Germany)**

[**Note:** The small circles represent the top class indicative of the largest city and the bottom class indicative of the small towns. The trend lines are based on the data points within the scaling ranges.]

# 4. Discussion

The empirical analysis shows that the natural cities of the four European and American countries follow hierarchical scaling laws. Taking scaling ranges into account, we can fit the size-number scaling and area-number scaling relations into the observational data very well. The two scaling relations are equivalent to the Zipf's law of distributions of urban population and area (Chen, 2012). The size-number scaling analysis gives the fractal dimension of population-size distribution $D$, and the area-number scaling analysis yields the fractal dimension of area-size distribution $d$. From the above-mentioned scaling relations, we can derive the size-area allometric scaling relation.

Hierarchical allometry is one of urban allometric scaling relations. Allometric scaling includes longitudinal allometry (temporal allometry), transversal allometry, and spatial allometry (Chen, 2008; Pumain and Moriconi-Ebrard, 1997), and transversal allometry includes cross-sectional allometry based on rank-size distribution and hierarchical allometry based on cascade structure. The longitudinal allometry is based on exponential growth (Bertalanffy, 1968), or logistic growth (Chen, 2014), while the transversal allometry is based on rank-size distribution, exponential distribution, or hierarchical power-law distribution (Table 4). The hierarchical allometry is equivalent in mathematics to the cross-sectional allometry, and form a connection between transversal allometry and longitudinal allometry. A hierarchy makes a link between spatial disaggregation and network structure (Batty and Longley, 1994). By researching allometric scaling in hierarchies of cities, we will be able to find the latent inherent correlations between spatial patterns, temporal processes, and dynamic mechanisms of urban evolution. The allometric scaling has been applied to urban studies based on census data and statistical data (Chen, 2011; Chen, 2012). However, the observational data of natural cities provide better evidences for the hierarchical allometric scaling laws.

Table 4 The longitudinal and transversal allometric scaling relations of cities and the related growth or distribution functions

| Type | Sub-type | Basic models | Main model | Parameters |
|------|----------|--------------|------------|------------|
| Longitudinal allometry | Exponential allometry | $S_t = S_0 e^{ut}$ <br> $A_t = A_0 e^{vt}$ | $A_t = a S_t^b$ | $a = A_0 S_0^{-b}$ <br> $b = v/u$ |
| | Logistic allometry | $S_t = \dfrac{S_{\max}}{1+(S_{\max}/S_0 -1)e^{-vt}}$ <br> $A_t = \dfrac{A_{\max}}{1+(A_{\max}/A_0 -1)e^{-ut}}$ | $\dfrac{A_t}{A_{\max}-A_t} = a\left(\dfrac{S_t}{S_{\max}-S_t}\right)^b$ | $a = \dfrac{A_0}{A_{\max}-A_0}$ <br> $\div \left(\dfrac{S_0}{S_{\max}-S_0}\right)^b$ <br> $b = v/u$ |
| Crosssectional allometry | Power allometry | $S_k = S_1 k^{-q}$ <br> $A_k = A_1 k^{-p}$ | $A_k = a S_k^b$ | $a = A_1 S_1^{-b}$ <br> $b = p/q$ |
| Hierarchical allometry | Exponential allometry | $S_m = S_1 r_s^{1-m}$ <br> $A_m = A_1 r_a^{1-m}$ | $A_m = a S_m^b$ | $a = A_1 S_1^{-b}$ <br> $b = \ln r_a / \ln r_s$ |
| | Power allometry | $S_m = S_1 N_m^{-q}$ <br> $A_m = A_1 N_m^{-p}$ | $A_m = a S_m^b$ | $a = A_1 S_1^{-b}$ <br> $b = p/q$ |

**Note**: The symbols are as follows: $t$—time; $k$—rank; $m$—level; $S$—(population) size; $A$—urban area; $a$, $b$, $p$, $q$, $u$, $v$, $r_a$, $r_p$, $A_0$, $A_1$, $A_{\max}$, $S_0$, $S_1$, $S_{\max}$ are all parameters (proportionality coefficient, scaling exponent, ratio, capacity, etc.).

Allometric scaling indicates a power-law relation, which suggests a proportional relation between two measures. Therefore, allometry is involved with two concepts of modern mathematical modeling. One is spatial dimension, and the other is scaling range. Spatial dimension is one of the conundrums in mathematical description. A measure is proportional to another measure, if and only if (iff) the two measures bear the same spatial dimension. So a length is in proportion to the square root of an area, or to the cubic root of a volume. This principle has long been discovered by the ancient Greeks. In this sense, the allometric scaling exponent of size-area indicates the ratio of two spatial dimension values such as $b=D/d=D_a/D_s$, where $D$ and $d$ refer to the fractal dimensions of population and area size distributions, and $D_s$ and $D_a$ denote the fractal dimension of the spatial distributions of city population size and land use form. In fact, $D$ and $d$ can be proved to be paradimension according to the relationships between Zipf's law and hierarchical scaling law. The concept of paradimension was sublated by Mandelbrot (1982), but it is useful in the studies on fractal cities. Scaling range indicates an effective range for scale-free analysis of cities. The hierarchical allometry is based on Zipf's distributions. The largest city and the very small town may violate Zipf's law and take on outliers on a double logarithmic plot. If the largest city (e.g., London) is a world city, and the area of its country (Great Britain) is not large, then the sphere of influence of the largest city will go far beyond the national area. As a result, the largest city becomes on an outlier and the primate distribution will replace Zipf's distribution of cities (Chen, 2008). Meanwhile, the small towns may form outliers and go beyond the scaling range in a log-log plot due to undergrowth of city sizes (Chen, 2012). In China, improper government intervention in urbanization often gives rise to abnormality of urban structure, which takes on outliers in datasets (Chen, 2014). Sometimes, small cities or towns are developed in relative size, but the city number does not reach $2^M$, where $M$ is a positive integer. Thus the last level of urban hierarchy forms a lame-duck class (Davis, 1978).

The power law behaviors of hierarchical scaling in city development can be explained by the principle of entropy maximization. A power law is based on two exponential laws, and the relationships between the power laws and exponential laws can be revealed by the self-similar hierarchy. In fact, exponential distributions can be derived by using entropy-maximizing methods (Bussière and Snickars, 1970; Chen, 2012; Curry, 1964; Wilson, 1968), and a power law can be derived from a pair of exponential laws (Chen, 2008; Chen, 2012; Wilson, 2010). In this paper, the power laws, equations (4), (5), and (6), are derived from the exponential laws, equations (1), (2), and (3). It has been proved that equations (1), (2), and (3) can be derived from entropy maximization principles of urban evolution (Chen, 2012). City size can be measured by both urban population and urbanized area. Equation (1) is based on the entropy maximization process of urban frequency distribution, while equations (2) and (3) are based on the entropy maximization of city size distributions (Table 5). This suggests that a power law is based on two dual processes of entropy maximization. Entropy maximization means an optimal and coordinated relationships between the efficiency of the whole and the equity among individuals in a self-organized system (Chen, 2012; Chen, 2015b). The results of empirical analyses indicates that the entropy maximization principle can be employed to explain the evolution of natural cities.

**Table 5 Two types of entropy maximization processes in the evolution of city size distributions**

| Entropy process | Law | Formula | Equation | Complexity |
|---|---|---|---|---|
| Entropy maximization of frequency distribution | Number law | $N_m = N_1 r_n^{m-1}$ | 1 | External complexity |

| Entropy maximization of size distribution | Population law | $S_m = S_1 r_s^{1-m}$ | 2 | Internal complexity |
|---|---|---|---|---|
| | Area law | $A_m = A_1 r_a^{1-m}$ | 3 | Internal complexity |

The merits of this study rest with data quality, dataset size, and mathematical models. On the one hand, all the observational data are based on the concept of natural cities and bear high quality. On the other, the size of datasets are very large compared the traditional sample sizes for rank-size analysis. Compared with the studies on urban hierarchies and rank-size distributions based on census data or statistical data, the datasets of natural cities are more suitable for hierarchical scaling analysis of cities. What is more, the models have performance of anti-disturbance of random noises. The main drawbacks of the work lie in two aspects. First, the city size is measured by numbers of block or traffic nodes rather than urban population. A city is a human settlement, and population size belongs to the first order dynamic models of cities (Arbesman, 2012). Two central variables can be employed to research spatial dynamics of urban development: population and wealth (Dendrinos, 1992). If the relation between urban population and block/node number is linear, the number of blocks or traffic nodes can be used to replace urban population, otherwise, the real relation should be revealed. Second, the temporal dimension does not be considered. Only one year datasets are available, and we cannot examine the dynamic change of hierarchies of natural cities. Despite these shortcomings, the contribution of the paper is clear: we use four big datasets of high quality to verify the hierarchical scaling laws from urban angle of view.

## 5. Conclusions

In this paper, we investigate the systems of natural cities in four European and American counties. Two measures are employed to reflect city size, one is the number of blocks, and the other is number of streets nodes. Different urban systems based on different size measurements lead to the same direction: all these systems of cities can be organized into hierarchies with cascade structure. The self-similar hierarchy can be described with a set of exponential laws: number law, population size law, and urban area law. The three exponential equations can be equivalently transformed into a set of power functions, the first one reflect the size-number scaling relation, the second one reflect the area-number scaling relation, and third one reflect the size-area allometric scaling relation. The self-similar hierarchy indicates a kind of deep structure of systems of cities and latent spatial order in urban evolution.

The main conclusions of this study can be drawn as follows. **First, the natural cities follow the hierarchical scaling laws, which can be represented by a set of power functions**. The three hierarchical models can be well fitted to the datasets of city size and urban area of America, Britain, France, and Germany by taking the scaling range into consideration. Compared with the census data or statistic data of cities, the observational data of natural cities show better effect of hierarchical scaling analysis. Moreover, the allometric scaling relation comes from a pair of rank-size scaling relations. In theory, the allometric scaling exponent is equal to the ratio of the fractal dimension of population size distribution to that of area size distribution; in practice, the allometric exponent is very close to the quotient of the two fractal dimension values of size distributions (esp. Britain and Germany). **Second, the principle of entropy maximization can be employed to explain the power law behaviors in hierarchies of natural cities.** A power law is based on a pair of exponential laws. An exponential law can be derived by means of the method of entropy maximizing. Thus a power law is determined by two entropy maximization processes. An urban hierarchy is involved two types of entropy maximization: frequency distribution and size distributions. The fractal models are controlled by an entropy maximization process of frequency distribution and that of size distribution, while an allometric scaling relation is dominated by two entropy maximization processes of size distributions. Entropy maximization can explain the power law of traditional city size distribution, but this principle seems to be more suitable for explaining the evolution and power law emergence of natural cities. **Third, the primate distribution of city sizes influences the hierarchical scaling patterns to some extent**. In urban geography, city size distributions are divided into two different groups: rank-size distribution and primate distribution. However, according to this study, the primate distribution seems not to represent an independent type. The large cities in Britain and France take on the character of primate distribution because London and Paris are two global cities. The primate distribution has impact on the log-log relations between city number, size, and urban area. However, this influence is not significant. This suggests that, compared with the rank-size law, the primate law represents a local rule, which can be covered by the hierarchical scaling relations based on big datasets.

**Acknowledgements**

This research was sponsored by the National Natural Science Foundations of China (Grant No. 41590843 & 41671167). The supports are gratefully acknowledged.

# References

Arbesman S (2012). *The Half-Life of Facts: Why Everything We Know Has An Expiration Date*. New York: Penguin Group

Arcaute E, Hatna E, Ferguson P, Youn H, Johansson A, Batty M (2015). Constructing cities, deconstructing scaling laws. *Journal of the Royal Society Interface*, 12(102): 20140745

Bak P (1996). *How Nature Works: The Science of Self-Organized Criticality*. Springer: New York

Batty M (2006). Hierarchy in cities and city systems. In: D. Pumain. *Hierarchy in Natural and Social Sciences*. Dordrecht: Springer, pp143-168

Batty M (2008). The size, scale, and shape of cities. *Science*, 319: 769-771

Batty M, Longley PA (1994). *Fractal cities: A geometry of form and function*. London: Academic Press

Beckmann MJ (1958). City hierarchies and distribution of city sizes. *Economic Development and Cultural Change*, 6(3): 243-248

Bertalanffy L von (1968). *General System Theory: Foundations, Development, and Applications*. New York: George Breziller

Bettencourt LMA (2013). The origins of scaling in cities. *Science*, 340: 1438-1441

Bettencourt LMA, Lobo J, Helbing D, Kühnert C, West GB (2007). Growth, innovation, scaling, and the pace of life in cities. *PNAS*, 104(17): 7301-7306

Bussière R, Snickars F (1970). Derivation of the negative exponential model by an entropy maximizing method. *Environment and Planning A*, 2(3): 295-301

Charnes A, Frome EL, Yu PL (1976). The equivalence of generalized least squares and maximum likelihood estimates in the exponential family. *Journal of the American Statistical Association*, 71 (353): 169–171

Chen YG (2008). *Fractal Urban Systems: Scaling, Symmetry, and Spatial Complexity*. Beijing: Science Press [In Chinese]

Chen YG (2010). Characterizing growth and form of fractal cities with allometric scaling exponents. *Discrete Dynamics in Nature and Society*, Volume 2010, Article ID 194715, 22 pages

Chen YG (2011). Fractal systems of central places based on intermittency of space-filling. *Chaos, Solitons & Fractals*, 44(8): 619-632

Chen YG (2012). The rank-size scaling law and entropy-maximizing principle. *Physica A: Statistical Mechanics and its Applications*, 391(3): 767-778

Chen YG (2014). An allometric scaling relation based on logistic growth of cities. *Chaos, Solitons & Fractals*, 65: 65-77

Chen YG (2015a). Fractals and fractal dimension of systems of blood vessels: An analogy between artery trees, river networks, and urban hierarchies. *Fractal Geometry and Nonlinear Analysis in Medicine and Biology*, 1(2):26-32

Chen YG (2015b). The distance-decay function of geographical gravity model: power law or exponential law? *Chaos, Solitons & Fractals*, 77: 174-189

Chen YG (2016). The evolution of Zipf's law indicative of city development. *Physica A: Statistical Mechanics and its Applications*, 443: 555-567

Christaller W (1933). *Central Places in Southern Germany* (translated by C. W. Baskin, 1966). Englewood Cliffs, NJ: Prentice Hall

Clauset A, Shalizi CR, Newman MEJ (2009). Power-law distributions in empirical data. *Siam Review*, 51(4): 661-703

Curry L (1964). The random spatial economy: an exploration in settlement theory. *Annals of the Association of American Geographers*, 54(1): 138-146

Davis K (1978). World urbanization: 1950-1970. In: *Systems of Cities* (ed. I.S. Bourne and J.W. Simons). New York: Oxford University Press, pp92-100

Dendrinos DS (1992). *The Dynamics of Cities: Ecological Determinism, Dualism and Chaos*. London and New York: Routledge

Frankhauser P (1994). *La Fractalité des Structures Urbaines (The Fractal Aspects of Urban Structures)*. Paris: Economica

Frankhauser P (1998). The fractal approach: A new tool for the spatial analysis of urban agglomerations. *Population: An English Selection*, 10(1): 205-240

Gabaix X (1999a). Zipf's law and the growth of cities. *The American Economic Review*, 89(2):

129-132

Gabaix X (1999b). Zipf's law for cities: an explanation. *Quarterly Journal of Economics*, 114 (3): 739–767

Goodchild MF, Mark DM (1987). The fractal nature of geographical phenomena. *Annals of Association of American Geographers*, 77(2): 265-278

Gutenberg B, Richter CF (1954). *Seismicity of the Earth and Associated Phenomenon* (2nd edition). Princeton: Princeton University Press

Horton RE (1945). Erosional development of streams and their drainage basins: Hydrophysical approach to quantitative morphology. *Bulletin of the Geophysical Society of America*, 56(3): 275-370

Jiang B (2013). The image of the city out of the underlying scaling of city artifacts or locations. *Annals of the Association of American Geographers*, 103(6): 1552–1566

Jiang B, Jia T (2011). Zipf's law for all the natural cities in the United States: a geospatial perspective. *International Journal of Geographical Information Science*, 25(8): 1269-1281

Jiang B, Liu XT (2012). Scaling of geographic space from the perspective of city and field blocks and using volunteered geographic information. *International Journal of Geographical Information Science*, 26(2): 215-229

Jiang B, Yao X (2010eds). *Geospatial Analysis and Modeling of Urban Structure and Dynamics*. Berlin: Springer

Jiang B, Yin J (2014). Ht-index for quantifying the fractal or scaling structure of geographic features. *Annals of the Association of American Geographers*, 104(3): 530-541

Jiang ZL, He GC (1989). Geometrical morphology of the human coronary arteries. *Journal of Third Military Medical University*, 11(2): 85-91 [In Chinese]

Jiang ZL, He GC (1990). Geometrical morphology of coronary arteries in dog. *Chinese Journal of Anatomy*, 13(3): 236-241 [In Chinese]

Lobo J, Bettencourt LMA, Strumsky D, West GB (2013). Urban scaling and the production function for cities. *PLoS ONE*, 8(3): e58407

Longley PA, Batty M, Shepherd J (1991). The size, shape and dimension of urban settlements. *Transactions of the Institute of British Geographers (New Series)*, 16(1): 75-94

Louf R, Barthelemy M (2014a). Scaling: lost in the smog. *Environment and Planning B: Planning and*

*Design*, 41: 767-769

Louf R, Barthelemy M (2014b). How congestion shapes cities: from mobility patterns to scaling. *Scientific Reports*, 4: 5561

Mandelbrot BB (1982). *The Fractal Geometry of Nature*. New York: W. H. Freeman and Company

Newman MEJ (2005). Power laws, Pareto distributions and Zipf's law. *Contemporary Physics*, 46(5): 323-351

Pumain D (2006 ed). *Hierarchy in Natural and Social Sciences*. Dordrecht: Springer

Pumain D, Moriconi-Ebrard F (1997). City size distributions and metropolisation. *GeoJournal*, 43 (4): 307-314

Rozenfeld HD, Rybski D, Andrade Jr. DS, Batty M, Stanley HE, Makse HA (2008). Laws of population growth. *PNAS*, 105(48): 18702-18707

Rozenfeld HD, Rybski D, Gabaix X, Makse HA (2011). The area and population of cities: New insights from a different perspective on cities. *American Economic Review*, 101(5): 2205–2225

Rybski D, Buldyrev SV, Havlin S, Liljeros F, Hernán A, Makse HA (2009). Scaling laws of human interaction activity. *PNAS*, 106(31): 12640–12645

Schumm SA (1956). Evolution of drainage systems and slopes in badlands at Perth Amboy, New Jersey. *Geological Society of America Bulletin*, 67(5): 597-646

Strahler AE (1952). Hypsometric (area-altitude) analysis of erosional topography. *Geological Society of American Bulletin*, 63(11): 1117-1142

Waldrop M (1992). *Complexity: The Emerging of Science at the Edge of Order and Chaos*. New York: Simon and Schuster

West GB, Woodruff WH, Brown JH (2002). Allometric scaling of metabolic rate from molecules and mitochondria to cells and mammals. *PNAS*, 99: 2473-2478

Williams GP (1997). *Chaos Theory Tamed*. Washington, D.C.: Joseph Henry Press

Wilson AG (1968). Modelling and systems analysis in urban planning. *Nature*, 220: 963-966

Wilson AG (2010). Entropy in urban and regional modelling: retrospect and prospect. *Geographical Analysis*, 42(4): 364-394

Woldenberg MJ, Berry BJL (1967). Rivers and central places: analogous systems? *Journal of Regional Science*, 7(2): 129-139

Zhang J, Yu TK (2010). Allometric scaling of countries. *Physica A: Statistical Mechanics and its*

*Applications*, 389(21): 4887-4896

Zipf GK (1949). *Human Behavior and the Principle of Least Effort*. Reading, MA: Addison–Wesley